# Computations with one and two real algebraic numbers

Ioannis Z. Emiris and Elias P. Tsigaridas
Department of Informatics and Telecommunications
National University of Athens, HELLAS
{emiris,et}@di.uoa.gr

### Abstract

We present algorithmic and complexity results concerning computations with one and two real algebraic numbers, as well as real solving of univariate polynomials and bivariate polynomial systems with integer coefficients using Sturm-Habicht sequences.

Our main results, in the univariate case, concern the problems of real root isolation (Th. 19) and simultaneous inequalities (Cor. 26) and in the bivariate, the problems of system real solving (Th. 42), sign evaluation (Th. 37) and simultaneous inequalities (Cor. 43).

## 1 Introduction

In what follows $\mathcal{O}_B$ means bit complexity and the $\widetilde{\mathcal{O}}_B$-notation means that we are ignoring logarithmic factors.

For a polynomial $P \in \mathbb{Z}[X, Y]$, $\deg(P)$ denotes its total degree while $\deg_X(P)$ (resp. $\deg_Y(P)$) denotes its degree if we consider it as a univariate polynomial with respect to $X$ (resp. $Y$).

By $\mathcal{L}(P)$ we denote an upper bound on the bit size of the coefficients of $P$ (including a bit for the sign) i.e $\mathcal{L}(P) = \lfloor \lg(\max |a_i|) \rfloor + 2$, where $a_i$ are the coefficients of $P$. In particular $\mathcal{L}(a)$ is the bit size of $a$, if it is a non zero integer, or the maximum bit size of the numerator and the denominator if it is a rational number.

Let $M(\tau)$ denote the bit complexity of multiplying two integers of bit size at most $\tau$ and $M(d, \tau)$ denote the bit complexity of multiplying two univariate polynomials of degrees bounded by $d$ and coefficient bit size at most $\tau$. Using fast multiplication algorithms, i.e FFT [5, 40, 42], the complexities of these operations are

$$\begin{aligned} M(\tau) &= \mathcal{O}(\tau \lg \tau \lg \lg \tau) \\ M(d, \tau) &= \mathcal{O}_B(d\tau \lg(d\tau) \lg \lg(d\tau)) \end{aligned} \tag{1}$$

Hence $M(\tau) = \mathcal{O}_B(\tau \lg^{c_1} \tau)$ and $M(d, \tau) = \mathcal{O}_B(d\tau \lg^{c_2}(d\tau))$ for suitable constants $c_1, c_2$.

## 2 Sturm–Habicht sequences

Let $A = \sum_{k=0}^{p} a_k X^k$, $B = \sum_{k=0}^{q} b_k X^k \in \mathbb{Z}[X]$ where $\deg(A) = p > q = \deg(B)$ and $\mathcal{L}(A) = \mathcal{L}(B) = \tau$.

We denote by $\texttt{rem}(A, B)$ and $\texttt{quo}(A, B)$ the remainder and the quotient, respectively, of the division of $A$ and $B$. In general both may have rational coefficients. Following [26] we give the following definitions:

**Definitions 1** *The signed polynomial remainder sequence of $A$ and $B$, $\textbf{sPRS}(A, B)$ is the polynomial sequence*

$$(R_0 = A, R_1 = B, R_2 = -\texttt{rem}(A, B), \dots, R_k = -\texttt{rem}(R_{k-2}, R_{k-1}))$$

*where $\texttt{rem}(R_{k-1}, R_k) = 0$.*

*The quotient sequence of* $A$ *and* $B$ *is the polynomial sequence*

$$(Q_0 = \mathrm{quo}\,(R_0, R_1)\,, Q_1 = \mathrm{quo}\,(R_1, R_2)\,, \ldots, Q_{k-1} = \mathrm{quo}\,(R_{k-1}, R_k))$$

*We also need the definition of the quotient boot, which is the polynomial sequence*

$$(Q_0, Q_1, \ldots, Q_{k-1}, R_k)$$

There is a huge bibliography on signed polynomial remainder sequences (c.f [2, 40, 42] and references there in). The subresultant techinques were introduced by Collins [8], in order to reduce the growth of the coefficient in the signed polynomial sequences when pseudo-division is used. Gathen and Lücking [41] presented a unified approach to various definitions and algorithms of the subresultants, while El Kahoui [16] studied the subresultants in arbitrary commutative rings. For the Sturm-Habicht (or Sylvester-Habicht) sequences the reader may refer to the work of Gonzalez-Vega et al [21, 22].

In this paper we consider the Sturm-Habicht sequence of $A$ and $B$, i.e **StHa**$(A, B)$, which contains polynomials that are proportional to the polynomials in **sPRS** $(A, B)$, i.e there exists an integer such that if we multiply a polynomial in **StHa**$(A, B)$ we find the corresponding polynomial in **sPRS** $(A, B)$. However Sturm-Habicht sequences achieve better bounds on the bit size of the coefficients and have good specialization properties, since they are defined through determinants. Moreover, they are obtained by sign modification of the subresultant sequence.

Let $M_j$ be the matrix which has as rows the coefficient vectors of the polynomials

$$AX^{q-1-j}, AX^{q-2-j}, \ldots, AX, A, B, BX, \ldots, BX^{p-2-j}, BX^{p-1-j}$$

with respect to the monomial basis $X^{p+q-1-j}, X^{p+q-2-j}, \ldots, X, 1$. The dimension of $M_j$ is $(p + q - 1 - 2j) \times (p + q - 1 - j)$.

For $l = 0, \ldots, p + q - 1 - j$ let $M_j^l$ be the square matrix of dimension $(p + q - 2j) \times (p + q - 2j)$ obtained by taking the first $p + q - 1 - 2j$ columns and the $l$-th column of $M_j$.

**Definition 2** *The Sturm-Habicht sequence of* $A$ *and* $B$, *is the sequence*

$$\mathbf{StHa}(A, B) = (H_p = H_p(A, B), \ldots, H_0 = H_o(A, B))$$

*where* $H_p = A$, $H_{p-1} = B$ *and* $H_j = \sum_{l=0}^{j} \det(M_j^l) X^l$.

*The sequence of principal Sturm-Habicht coefficients*

$$(h_p = h_p(A, B), \ldots, h_0(A, B))$$

*is defined as* $h_p = 1$ *and* $h_j = \mathtt{coeff}_j(H_j)$ *is the coefficient of* $X^j$ *in the polynomial* $H_j$ *for* $0 \le i \le p$.
*When* $h_j = 0$ *for some* $j$ *then the sequence is called defective, otherwise non-defective.*

If the **StHa**$(A, B)$ is non-defective then it coincides up to sign with the classical subresultant sequence. The sign of proportionality is $(-1)^{\frac{(p-j)(p-j-1)}{2}}$. However, in the defective case, one has better control on the bit size of the coefficients in the sequence.

One important property [2] is that the polynomial $H_0(A, B)$, modulo its sign, is the resultant of $A$ and $B$. Moreover the greatest common divisor of $A$ and $B$ is obtained as a by-product, together with the following equivalence:

$$H_k(A, B) = \gcd(A, B) \Leftrightarrow \left\{ \begin{array}{c} h_0(A, B) = \cdots = h_{k-1}(A, B) = 0 \\ h_k(A, B) \ne 0 \end{array} \right.$$

There are various algorithms that compute all polynomials in **StHa**$(A, B)$, e.g [2, 14, 27, 40, 42]. Most algorithms exploit the special structure of the cofficients of the polynomials that appear in the sequence and manage to reduce their bit size. All of these algorithms have more or less similar arithmetic and bit complexity.

**Theorem 3** *[2, 33, 14, 27] There is an algorithm that computes* $\mathbf{StHa}(A, B)$ *in* $\mathcal{O}_B(pq\,M(p\tau))$, *or* $\widetilde{\mathcal{O}}_B(p^2q\tau)$. *Moreover,* $\mathcal{L}(H_j(A, B)) = \mathcal{O}(p\tau)$.

The complexity for the computation of the whole $\mathbf{StHa}(A, B)$ is optimal up to constants and some logarithmic factors. Notice that there are $\Omega(q)$ polynomials in the sequence which have degree $\Omega(p)$, hence the total number of all the coefficients appeared in the sequence is $\Omega(pq)$. This observation allows us to argue that the arithmetic complexity $O(pq)$ achieved by the algorithms is optimal and since the bit size of the coefficients is $\Omega(p\tau)$ one can also, trivially, deduce the optimality of the bit complexity. However, there are cases where only one polynomial in the sequence is needed (e.g some polynomial in the middle of the sequence or the gcd, or the resultant etc) or we do not need the actual sequence but the evaluation of it over a number. In these cases a faster algorithm exists, which represents the sequence implicitly by the quotient boot. and is based on a divide-and-conquer strategy and on the idea of the HALF-GCD algorithm. The idea of this "implicit" representation is not new. The reader may refer to Strassen [39], where the optimality of this evaluation scheme is proven basen on the work of Knuth and Shönhage and to Yap [42] for the analysis of the HALF-GCD algorithm. Schwartz and Sharir [38] mentioned the benefits of this approach for computations with real algebraic numbers and also Davenport [11] exploited the advantages of this approach for real root isolation. Lickteig and Roy [26] and independently Reischert [33] formulated this approach for Sturm-Habicht sequences.

**Theorem 4** *[2, 26, 33, 40] The quotient boot, the resultant and the gcd of* $A$ *and* $B$, *can be computed in* $\mathcal{O}_B(q\lg q M(p\tau))$ *or* $\widetilde{\mathcal{O}}_B(p\,q\,\tau)$.

Actually for the computation of $\gcd(A, B)$ various algorithms exist with complexity $\widetilde{\mathcal{O}}_B(pq\tau)$ (c.f [40, 42]).

Let the quotient boot that corresponds to $\mathbf{StHa}(A, B)$, be $\mathbf{StHaQ}(A, B) = (Q_0, Q_1, \ldots, Q_{k-1}, H_k)$. The number of coefficients in $\mathbf{StHaQ}(A, B)$ is $\mathcal{O}(q)$ and their bit size is $\mathcal{O}(p\tau)$ (c.f [2, 33]). The evaluation of the sequence on a number can be recovered from the quotient boot starting from $H_k$.

**Theorem 5** *[26, 33] There is an algorithm that computes the evaluation of* $\mathbf{StHa}(A, B)$ *over a number* $a$, *where* $a \in \mathbb{Q} \cup \{\pm\infty\}$ *and has bit size at most* $\sigma$ *in* $\mathcal{O}_B(q\lg q M(\max(p\tau, q\sigma)))$ *or in* $\mathcal{O}_B(qM(\max(p\tau, q\sigma)))$ *if* $\mathbf{StHaQ}(A, B)$ *is already computed.*

*In both cases the complexity is* $\widetilde{\mathcal{O}}_B(q\max(p\tau, q\sigma))$.

**Remark 6** *In many cases (e.g real root isolation, comparison of algebraic numbers etc) we need the evaluation of* $\mathbf{StHa}(A, A^{'})$ *over a rational number of bit size* $\mathcal{O}(p\tau)$. *If we perform the evaluation by Horner's rule then since there are* $\Omega(p^2)$ *coefficients in the sequence and we must multiply numbers of bit size* $\mathcal{O}(p\tau)$ *and* $\mathcal{O}(p^2\tau)$, *the overall complexity is* $\mathcal{O}_B(p^3M(p\tau))$.

*However, when we compute the complete* $\mathbf{StHa}(A, A^{'})$ *in* $\mathcal{O}_B(p^2M(p\tau))$ *(Th. 3), the quotient boot is computed implicitly [33, 2]. Thus, we can use the quotient boot in order to perform the evaluation even if we have already computed all the polynomials in the Sturm-Habicht sequence.*

**Remark 7** *Notice also that the computation should be started by the last element of the quotient boot so as to avoid the costly computation of two polynomial evaluations using the Horner's scheme.*

*Even though this approach is optimal, we have to mention that involves big constants in its complexity, thus it is not efficient in practice if the length of the sequence is not sufficient big and special techniques should be used for its implementation in order to avoid costly operations with rational numbers.*

*So, as it is always the case with optimal algebraic algorithms, the implementation is far from a trivial task.*

We can also use Sturm-Habicht sequences in order to compute the square-free part of $A$, i.e $A_{\mathtt{red}} = A/\gcd(A, A^{'})$, where $A^{'}$ is the derivative of $A$.

**Theorem 8** *[2, Algorithm 10.17, p. 326] The square-free part of $A$, i.e. $A_{red}$, can be computed from $\mathbf{StHa}(A, A')$, in $\mathcal{O}_B(p \lg p M(p\tau))$ or $\widetilde{\mathcal{O}}_B(p^2\tau)$. Moreover, $\mathcal{L}(A_{red}) = \mathcal{O}(p + \tau)$.*

**Remark 9** *There is a normalization step [2] at the last element of $\mathbf{StHa}(A, A')$ so as to compute $A_{red}$ which achieves the good bound for $\mathcal{L}(A_{red})$. Notice that if we rely on Mignotte's bound [28] that applies to all exact polynomial divisors of $A$ or on the subresultant algorithm [42] then the bit size of $A_{red}$ is $\mathcal{O}(p\tau)$.*

Let $W_{(A,B)}(a)$ denote the number of modified sign changes of the evaluation of $\mathbf{StHa}(A, B)$ over $a$. Notice that $W_{(A,B)}(a)$ does not refer to the usual counting of sign varations, since special care should be taken for the presence of consecutive zeros [2, 22, 23].

**Theorem 10** *[2, 42, 34] Let $A, B \in \mathbb{Z}[X]$ be relatively prime polynomials, where is $A$ square-free and $A'$ is the derivative of $A$. If $a < b$ are both non-roots of $A$ and $\gamma$ ranges over the roots of $A$ in $(a, b)$, then*

$$W_{(A,B)}([a, b]) := W_{(A,B)}(a) - W_{(A,B)}(b) = \sum_\gamma \operatorname{sign}(A'(\gamma)B(\gamma)).$$

**Corollary 11** *[2, 42] If $B = A'$ then $\mathbf{StHa}(A, A')$ is the Sturm sequence and Th. 10 counts the number of real roots of $A$ in $(a, b)$.*

**Remark 12** *Actually Th. 10 computes the Cauchy index of the rational function $B/A$ [2].*

# 3 Real algebraic numbers

## 3.1 Univariate real root isolation

Let $f = \sum_{i=0}^d a_i X^i \in \mathbb{Z}[X]$, with $\deg(f) = d$ and $\mathcal{L}(f) = \tau$ and let $f_{red}$ be its square free part. We want to isolate the real roots of $f$, i.e to compute intervals with rational endpoints that contain one and only one root of $f$, as well as the multiplicity of every real root.

Various algorithms exist for polynomial real root isolation, but most of them focus on square-free polynomials. The interested reader may refer to the algorithm of Collins and Loos [9], where the real roots of the derivative are used in order to isolate the real roots of the polynomial, with complexity $\widetilde{\mathcal{O}}_B(d^9 + d^6\tau^3)$, to the work of Akritas [1] for an algorithm based on continued fractions, or to the work of Rouillier and Zimmermann [35] (and references therein) where a unified approach with optimal memory management is presented for various algorithms that depend on Descartes' rule of sign. The complexity of all the algorithms is no better than $\widetilde{\mathcal{O}}_B(d^6\tau^2)$. Moreover Eigenwillig et al [15] recently presented an algorithm for polynomials with bit stream coefficients which is based on Descartes' rule of sign and the properties of the Bernstein basis. The complexity of their randomized algorithm is $\mathcal{O}_B(d^4(\log(s\mathsf{ep}) + \tau)^2)$, where $s\mathsf{ep}$ is the separation bound (see Rem. 20), that is $\mathcal{O}(2^{d\tau})$, hence the complexity is $\mathcal{O}_B(d^6\tau^2)$.

However, we have to mention that the stated references are only the tip of the iceberg of the existing bibliography.

If we restrict ourselves to real root isolation using Sturm (or Sturm-Habicht) sequences the first complete complexity analysis is probably due to Collins and Loos [9], that state a complexity of $\widetilde{\mathcal{O}}_B(d^7\tau^3)$. Davenport [11] improves this bound to $\widetilde{\mathcal{O}}_B(d^4\tau^2)$ but does not present a formal proof that this bound holds for non square-free polynomials (Th. 8) and does not compute the multiplicities of the roots. Also Schwartz and Sharir [38] implicitly state this bound, but without a proof. Recently Du et.al [13] prove this bound for non square-free polynomials by giving a ingenious amortized-like argument for the number of subdivisions that must be performed.

In this section we will prove that we can isolate the real roots of a non square-free polynomial in $\widetilde{\mathcal{O}}_B(d^4\tau^2)$ and that in the same time we can also compute the multiplicities of the real roots.

**Remark 13** *We are not mentioning neither the algorithm of Mourrain et al [32], that is roughly speaking based on a combination of Descartes' rule and on the properties of Bernstein basis, nor its improvements [31], since currently we are working with B. Mourrain on obtaining a complexity bound for this algorithm similar to the bound of the algorithm that uses Sturm-Habicht sequences.*

**Algorithm 1 (Real Root Isolation using Sturm-Habicht Sequences)**

> `Input:` *A polynomial* $\mathtt{f} \in \mathbb{Z}[X]$, *with* $\deg \mathtt{f} = \mathtt{d}$ *and* $\mathcal{L}(\mathtt{f}) = \tau$.
> `Output:` *A list of intervals with rational endpoints, which contain one and only one root of* $\mathtt{f}$ *and the multiplicity over every real root.*

1. *Compute the square free part of* $\mathtt{f}$, *i.e* $\mathtt{f_{red}}$

2. *Compute* $\mathbf{StHa}(\mathtt{f_{red}})$

3. *Compute an interval* $\mathtt{I_0} = (-\mathtt{B}, \mathtt{B})$ *with rational endpoints that contains all the real roots. Initialize a queue* $\mathtt{Q}$ *with* $\mathtt{I_0}$.

4. *While* $\mathtt{Q}$ *is not empty do*

   > *Pop an interval* $\mathtt{I}$ *from* $\mathtt{Q}$ *and compute using Cor. 11 the number of roots in* $\mathtt{I}$.
   >
   > *If* $\mathtt{I}$ *contains no real roots, discard* $\mathtt{I}$.
   >
   > *If* $\mathtt{I}$ *contains one real root, output* $\mathtt{I}$.
   >
   > *If* $\mathtt{I}$ *contains more than one real root split it to* $\mathtt{I_L}$ *and* $\mathtt{I_R}$ *and push them to* $\mathtt{Q}$.

5. *Determine the multiplicities of the real roots, using the square-free factorization of* $\mathtt{f}$.

**Remark 14** *For a detailed description of Steps 2-4, that exploits the implementation details, the reader may refer to Davenport et al [12] or the more recent work of Du et al [13]. Notice that special care should be taken for the case that the middle of a tested interval is a root of the polynomial, which is a non trivial implementation issue.*

### 3.1.1 Complexity analysis of real root isolation

**Step 1** The computation of $\mathtt{f_{red}}$ can be done in $\widetilde{\mathcal{O}}_B(\mathtt{d}^2\tau)$ (Th. 8). Notice that $\mathcal{L}(\mathtt{f_{red}}) = \mathcal{O}(\mathtt{d} + \tau)$. We assume that $\mathtt{d} = \mathcal{O}(\tau)$, thus $\mathcal{L}(\mathtt{f_{red}}) = \mathcal{O}(\tau)$.

**Step 2** We do no need the complete sequence (Remark 6), we only need the quotient boot, thus this computation can be done in $\widetilde{\mathcal{O}}_B(\mathtt{d}^2\tau)$ (Th. 4). However, we may also assume that the complete sequence is computed, with complexity $\widetilde{\mathcal{O}}_B(\mathtt{d}^3\tau)$ (Th. 3), since this step is not the bottleneck of the algorithm.

**Step 3** The Cauchy bound [2, 40, 42] states that if $\alpha$ is a real root of $\mathtt{f}$ then $|\alpha| \leq \mathtt{B} = 1 + \max\left(\left|\frac{a_{d-1}}{a_n}\right|, \left|\frac{a_{d-2}}{a_n}\right|, \ldots, \left|\frac{a_0}{a_n}\right|\right)$. Various upper bounds are known for the absolute value of the real roots (c.f [1, 2, 42, 40]). However, asymptotically the bit size of all the bounds is the same, i.e $\mathtt{B} \leq 2^\tau$.

**Step 4** We count the number of real roots using Cor. 11. by evaluating $\mathbf{StHa}(\mathtt{f_{red}})$ over rational numbers of bit size at most $\mathcal{O}(\mathtt{d}\tau)$ (Remark 20). The cost of every such evaluation is $\widetilde{\mathcal{O}}_B(\mathtt{d}^3\tau)$ (Th. 5). Since the number of subdivisions that we must perform in order to isolate all the real roots is $\mathcal{O}(\mathtt{d}\tau + \mathtt{d}\lg\mathtt{d})$ (Prop. 17), the overall complexity of this step is $\widetilde{\mathcal{O}}_B(\mathtt{d}^4\tau^2)$.

> Notice that the complexity of this step dominates the complexities of all the other steps.

**Step 5** In order to compute the multiplicities we compute the square-free factorization, i.e a sequence of square-free coprime polynomials $(g_1, g_2, \ldots, g_m)$ with $f = g_1 g_2^2 \cdots g_m^m$ and $g_m \neq 1$. The algorithm of Yun [40] computes the square free factorization in $\widetilde{\mathcal{O}}_B(d^2\tau)$. To be more specific the cost is twice the cost of the computation of **StHa**$(f, f')$ [19].

At every isolating interval, one and only one $g_j$ must have opposite signs at its endpoints, since $g_j$ are square free and pairwise coprime. If $g_j$ changes sign at an interval then the multiplicity of the real root that the interval contains is $j$.

We can evaluate each $g_i, 1 \leq i \leq m$, at all the isolating points simultaneously in $\widetilde{\mathcal{O}}_B(d^3\tau)$ [40, 42]. To be more precise we can perform an evaluation of a polynomial at $d$ numbers at the cost of evaluating the polynomial over one number using Horner's scheme. Since $m$ is at most $d$, the overall cost is $\widetilde{\mathcal{O}}_B(d^4\tau)$.

This bound is quite pessimistic due to the reason that it is not possible for the square free factorization of $f$ to contain $\mathcal{O}(d)$ polynomials of degree $\mathcal{O}(d)$ since this would lead to bound of $\mathcal{O}(d^2)$ for the degree of $f$.

Thus we may assume that either $m$ is a constant, or that the degrees of $g_j$'s are bounded by a constant. Hence the complexity of this step is $\widetilde{\mathcal{O}}_B(d^3\tau)$. However there is no need for a detailed study of the complexity of this step since both mentioned complexities do not dominate the complexity of the overall algorithm.

To complete the proof of the complexity of the algorithm we must prove that the number of subdivisions is $\mathcal{O}(d\tau + d \lg d)$. Recall [2] that the Mahler's measure, of $f$ is $\mathcal{M}(f) = |a_p| \prod_{i=1}^{d} \max\{1, |\gamma_i|\}$, where $a_p$ is the leading coefficient and $\gamma_i$ are all the roots of $f$. We know that $\mathcal{M}(f) < 2^\tau \sqrt{d+1}$ [2].

**Remark 15** *Since the product of the absolute values of roots greater than 1 of $f_{red}$ can not be greater than the one of $f$, we have the following inequality [2]*

$$\mathcal{M}(f_{red}) \leq \mathcal{M}(f) < 2^\tau \sqrt{d+1}$$

For the minimum distance between consecutive real roots of a square free polynomial the Davenport-Mahler bound is known [11]. The conditions of this bound where generalized by Du et al [13]. However, using Remark 15 we can provide a similar bound for non square free polynomials [11].

**Theorem 16 (Davenport-Mahler bound revisited)** *Let $\alpha_1 < \alpha_2 < \cdots < \alpha_k < \alpha_{k+1}$ be the $k+1$ distinct real roots of $f$, which is not necessarily square free, with $k \geq 1$. Then*

$$\mathcal{M}(f) \geq \prod_{i=1}^{k} |\alpha_i - \alpha_{i+1}| \geq \mathcal{M}(f)^{-d+1} d^{-\frac{d}{2}} (\frac{\sqrt{3}}{d})^k$$

**Proof**. If $f$ is square free then the bounds hold [11].

If it is not square free let $n \leq d$ be the degree of $f_{red}$. Notice that $k + 1 \leq n < d$. Since the bound holds for $f_{red}$, we have

$$
\begin{array}{rclcll}
\mathcal{M}(f_{red}) & \geq & \prod_{i=1}^{k} |\alpha_i - \alpha_{i+1}| & \geq & \mathcal{M}(f_{red})^{-n+1} n^{-\frac{n}{2}} (\frac{\sqrt{3}}{n})^k & \\
\mathcal{M}(f) & \geq & \prod_{i=1}^{k} |\alpha_i - \alpha_{i+1}| & \geq & \mathcal{M}(f)^{-n+1} n^{-\frac{n}{2}} (\frac{\sqrt{3}}{n})^k & \text{(Rem. 15)} \\
\mathcal{M}(f) & \geq & \prod_{i=1}^{k} |\alpha_i - \alpha_{i+1}| & \geq & \mathcal{M}(f)^{-d+1} d^{-\frac{d}{2}} (\frac{\sqrt{3}}{d})^k & \text{($n \leq d$)}
\end{array}
$$

Thus the theorem holds. $\qquad\qquad\square$

**Proposition 17** *The number of subdivisions that we need to perform in order to isolate the real roots of $f_{red}$ using Sturm-Habicht sequences is at most $\mathcal{O}(d\tau + d \lg d)$.*

**Proof.** The proof follows [11] and uses the result of th. 16.

Let $\alpha_i$ be the real roots of $f_{red}$ in increasing order. We know that the roots are contained in an interval $I = (-B, B)$ (we can compute such an interval using for example the bound of Cauchy, i.e $B \leq 2^\tau$ [1, 2, 40]). If $f_{red}$ has $k + 1$ real roots then we need to find $k$ separation points. For two consecutive real roots $\alpha_i$ and $\alpha_{i+1}$ a separation point for them is of magnitude $\frac{1}{2}|\alpha_i - \alpha_{i+1}|$ and we can compute it with at most $\lceil \lg \frac{2B}{|\alpha_i - \alpha_{i+1}|} \rceil$ subdivisions, using binary search.

Let $\mathcal{S}(I)$ denote the total number of subdivisions that we need to perform in order to compute all the isolating points of $f_{red}$ in $I$. Then

$$\mathcal{S}(I) \leq \sum_{i=1}^{k} \left\lceil \lg \frac{2B}{|\alpha_i - \alpha_{i+1}|} \right\rceil < k + k \lg(2B) - \sum_{i=1}^{k} \lg|\alpha_i - \alpha_{i+1}|$$

where the additional $k$ in the last inequality represents the $k$ possible roundings.

$$
\begin{aligned}
\mathcal{S}(I) \quad &< \quad k + k\lg(2B) + (d-1)\lg \mathcal{M}(f) + \tfrac{d}{2}\lg d - k\lg\sqrt{3} + k\lg d & \text{(Th. 16)} \\
&\leq \quad k + k\lg(2B) + (d-1)\lg(2^\tau\sqrt{d+1}) + \tfrac{d}{2}\lg d - k\lg\sqrt{3} + k\lg d & \text{(Rem. 15)} \\
&\leq \quad k + k(\tau+1) + (d-1)\tau + 2d\lg d + k\lg d & (B = 2^\tau)
\end{aligned}
$$

Furthermore, if $k \leq d - 1 \Rightarrow k < d$ then $\mathcal{S}(I) = \mathcal{O}(d\tau + d\lg d)$. □

**Remark 18** *Du et al [13] obtained a similar bound using a charging scheme (amortized analysis) for each subdivision that may provide better constants and that can also be used for complex root isolation.*

Under the hypothesis that $d = \mathcal{O}(\tau)$ (to simplify notation), the previous analysis leads to the following theorem

**Theorem 19** *Let $f \in \mathbb{Z}[X]$, with $\deg(f) = d$ and $\mathcal{L}(f) = \tau$, not necessarily square-free. We can isolate the real roots of $f$ and determine their multiplicities using Sturm-Habicht sequences in $\widetilde{\mathcal{O}}_B(d^4\tau^2)$. Moreover, the numerator and the denominator of the endpoints of the isolating intervals have bit size bounded by $\mathcal{O}(d\,\tau)$.*

**Remark 20** *The bit size of the endpoints of the isolating intervals is bounded by the bit size of the separation bound of $f$, i.e the minimum distance between two consecutive real roots. It is known [2, 28, 42] that $\operatorname{sep}(f) = \operatorname{sep}(f_{red}) \geq d^{-\frac{d+2}{2}}(d+1)^{\frac{1-d}{2}}2^{\tau(1-d)}$, thus $\lg(\operatorname{sep}(f)) = \lg(\operatorname{sep}(f_{red})) = \mathcal{O}(d\tau)$.*

*If a separation bound only on the real roots is needed then a bound due to Rump [43] may be used, i.e $\operatorname{sep}(f) > d^{-\frac{d+2}{2}}\sqrt{8}\left(1 + \max|a_i|^d\right)^{-1}$, which is a bit sharper but asymptotically has the same bit size.*

**Remark 21** *An $\widetilde{\mathcal{O}}_B(d^6(\tau + \lg d)^2)$ bound on the complexity for univariate real root isolation can be obtained if we adopt the algorithm of Mourrain et al [32], which is based on Bernstein basis and seems to have the best complexity in practice. The interested reader may refer to [2, 32] for more details.*

## 3.2   Representation of real algebraic numbers

The real algebraic numbers, i.e. those real numbers that satisfy a polynomial equation with integer coefficients, form a real closed field denoted by $\mathbb{R}_{alg} = \overline{\mathbb{Q}}$. From all integer polynomials that have an algebraic number $\alpha$ as root, the one with the minimum degree is called *minimal polynomial*. The minimal polynomial is unique, primitive and irreducible [12, 42]. In our approach, since we use Sturm-Habicht sequences, it suffices to deal with algebraic numbers, as roots of any square-free polynomial and not as roots of their minimal ones.

In order to represent a real algebraic number we chose the *isolating interval representation*.

**Definition 22** *The isolating-interval representation of real algebraic number* $\alpha \in \mathbb{R}_{\mathtt{alg}}$ *is* $\alpha \cong (\mathsf{P}(\mathsf{X}), \mathsf{I})$, *where* $\mathsf{P}(\mathsf{X}) \in \mathbb{Z}[\mathsf{X}]$ *is square-free and* $\mathsf{P}(\alpha) = 0$, $\mathsf{I} = [\mathtt{a}, \mathtt{b}]$, $\mathtt{a}, \mathtt{b}, \in \mathbb{Q}$ *and* $\mathsf{P}$ *has no other root in* $\mathsf{I}$.

**Remark 23** *Using the results of Section 3.1, we conclude that we can represent all the roots of a polynomial* $\mathsf{f}$*, with* $\deg(\mathsf{f}) = \mathtt{d}$ *and* $\mathcal{L}(\mathsf{f}) = \tau$*, using isolating interval representation in* $\widetilde{\mathcal{O}}_{\mathsf{B}}(\mathtt{d}^4\tau^2)$ *and that the endpoints of the isolating intervals have bit size* $\mathcal{O}(\mathtt{d}\tau)$.

*This will be the case for all the real algebraic numbers that we will consider for the rest of the paper, except stated otherwise.*

## 3.3 Comparison and sign evaluation

We can use Sturm-Habicht sequences in order to find the sign of a univariate polynomial, evaluated over a real algebraic number and to compare two algebraic numbers (cf. [17] for degree $\leq 4$, where it is proven that these operations can be performed in $\mathcal{O}(1)$), or $\mathcal{O}_{\mathsf{B}}(\tau)$).

**Lemma 24** *[17, 34, 42] Let* $\mathsf{Q}(\mathsf{X}) \in \mathbb{Z}[\mathsf{X}]$*, where* $\deg(\mathsf{Q}) = \mathtt{d}$ *and* $\mathcal{L}(\mathsf{Q}) = \tau$*, and a real algebraic number* $\alpha \cong (\mathsf{P}, [\mathtt{a}, \mathtt{b}])$*. We can compute* $\mathrm{sign}(\mathsf{Q}(\alpha))$ *in* $\widetilde{\mathcal{O}}_{\mathsf{B}}(\mathtt{d}^3\tau)$.

**Proof.** By th. 10, $\mathrm{sign}(\mathsf{Q}(\alpha)) = \mathrm{sign}(\mathsf{W}_{\mathsf{P}, \mathsf{Q}}[\mathtt{a}, \mathtt{b}] \cdot \mathsf{P}'(\alpha))$. Thus we need to perforce two evaluations of $\mathbf{StHa}(\mathsf{P}, \mathsf{Q})$ over the endpoints of the isolating interval of $\alpha$. The complexity of each is $\widetilde{\mathcal{O}}_{\mathsf{B}}(\mathtt{d}^3\tau)$ (Th. 5 and Rem. 6), which is also the complexity of the operation. □

**Lemma 25** *[17, 42] We can compare two real algebraic numbers in isolating interval representation in* $\widetilde{\mathcal{O}}_{\mathsf{B}}(\mathtt{d}^3\tau)$.

**Proof.** Let two algebraic numbers $\gamma_1 \cong (\mathsf{P}_1(\mathsf{x}), \mathsf{I}_1)$ and $\gamma_2 \cong (\mathsf{P}_2(\mathsf{x}), \mathsf{I}_2)$ where $\mathsf{I}_1 = [\mathtt{a}_1, \mathtt{b}_1]$, $\mathsf{I}_2 = [\mathtt{a}_2, \mathtt{b}_2]$. Let $\mathsf{J} = \mathsf{I}_1 \cap \mathsf{I}_2$. When $\mathsf{J} = \emptyset$, or only one of $\gamma_1$ and $\gamma_2$ belong to $\mathsf{J}$, we can easily order the 2 algebraic numbers. If $\gamma_1, \gamma_2 \in \mathsf{J}$, then $\gamma_1 \geq \gamma_2 \Leftrightarrow \mathsf{P}_2(\gamma_1) \cdot \mathsf{P}_2'(\gamma_2) \geq 0$.

We obtain the sign of $\mathsf{P}_2'(\gamma_2)$, using Lem. 24, thus the complexity of comparison is $\widetilde{\mathcal{O}}_{\mathsf{B}}(\mathtt{d}^3\tau)$. □

## 3.4 Simultaneous inequalities

Let $\mathsf{P}, \mathsf{A}_1, \ldots, \mathsf{A}_{\mathtt{n}_1}, \mathsf{B}_1, \ldots, \mathsf{B}_{\mathtt{n}_2}, \mathsf{C}_1, \ldots, \mathsf{C}_{\mathtt{n}_3} \in \mathbb{Z}[\mathsf{X}]$, with degree bounded by $\mathtt{d}$ and coefficient bit size bounded by $\tau$. We wish to compute the number and the real roots, $\gamma$, of $\mathsf{P}$ such that $\mathsf{A}_{\mathtt{i}}(\gamma) > 0$, $\mathsf{B}_{\mathtt{j}}(\gamma) < 0$ and $\mathsf{C}_{\mathtt{k}}(\gamma) = 0$ and $1 \leq \mathtt{i} \leq \mathtt{n}_1, 1 \leq \mathtt{j} \leq \mathtt{n}_2, 1 \leq \mathtt{k} \leq \mathtt{n}_3$. Let $\mathtt{n} = \mathtt{n}_1 + \mathtt{n}_2 + \mathtt{n}_3$.

**Corollary 26** *There is an algorithm that solves the problem of simultaneous inequalities time* $\widetilde{\mathcal{O}}_{\mathsf{B}}(\max\{\mathtt{n}\,\mathtt{d}^4\tau, \mathtt{d}^4\tau^2\}) = \widetilde{\mathcal{O}}_{\mathsf{B}}(\mathtt{d}^4\tau\max\{\mathtt{n}, \tau\})$.

**Proof.** First we compute the isolating interval representation of all the real roots of $\mathsf{P}$ in $\widetilde{\mathcal{O}}_{\mathsf{B}}(\mathtt{d}^4\tau^2)$ (Th. 19). There are at most $\mathtt{d}$. For every real root $\gamma$ of $\mathsf{P}$, for every polynomial $\mathsf{A}_{\mathtt{i}}, \mathsf{B}_{\mathtt{j}}, \mathsf{C}_{\mathtt{k}}$ we compute the sign $(\mathsf{A}_{\mathtt{i}}(\gamma))$, $\mathrm{sign}\,(\mathsf{B}_{\mathtt{j}}(\gamma))$ and $\mathrm{sign}\,(\mathsf{C}_{\mathtt{k}}(\gamma))$.

Sign determination costs $\widetilde{\mathcal{O}}_{\mathsf{B}}(\mathtt{d}^3\tau)$ (Lem. 24) and in the worst case we must compute $\mathtt{n}$ of them. Thus the overall cost is $\widetilde{\mathcal{O}}_{\mathsf{B}}(\max\{\mathtt{n}\mathtt{d}^4\tau, \mathtt{d}^4\tau^2\})$. □

**Remark 27** *Ben-Or, Kozen and Reif [3] considered the problem of simultaneous inequalities. The interested reader may also refer to the work of Canny [6] for a variant that is faster in the univariate case (it saves a factor) which has arithmetic complexity* $\mathcal{O}(\mathtt{n}(\mathtt{m}\mathtt{d}\lg\mathtt{m}\lg^2\mathtt{d} + \mathtt{m}^{2.376}))$*, where* $\mathtt{m}$ *is the number of real roots of* $\mathsf{P}$.

*Coste and Roy [10] introduced Thom's encoding for the real roots of a polynomial and the problem of simultaneous inequalities in this encoding. Their approach is purely symbolic and works over arbitrary real closed fields. They also derive [36] polynomial bounds for the arithmetic complexity of the problem. This is not the case for our approach, which is pseudo-polynomial as all the approaches that depend on separation bounds, since for the real root isolation the number of subdivions that we must perform depends on the bit size of the coefficients. They state a complexity of $\widetilde{\mathcal{O}}_B(m^8)$, using fast multiplication algorithms but not fast computations and evaluation of polynomial sequences, where $m$ is an integer bigger that $n, d, \tau$. In this notation our bound is $\widetilde{\mathcal{O}}_B(m^6)$.*

*We are planning to perform a more thorough analysis of algorithms that use Thom's encoding by embedding the results of fast multiplication and fast computation and evaluation of Sturm-Habicht sequences, as well as an efficient implementation.*

**Remark 28** *In the book of Basu, Pollack and Roy [2] an algorithm for the problem is presented when the real algebraic numbers are in isolating interval representation, with complexity $\widetilde{\mathcal{O}}_B(nd^6\tau^2)$. However their analysis does not assume fast multiplication algorithms and they determine the sign of a polynomial over an algebraic number by repeated refinements of the isolating intervals.*

### 3.5 Computation in an extension field

If $\alpha \cong (A, I)$, then we can perform computations in $\mathbb{Q}(\alpha)$. If $\deg(A) = d$ then $\mathbb{Q}(\alpha)$ is a vector space with basis elements $(1, \alpha, \ldots, \alpha^{d-1})$ [42]. Thus we can represent every element $\beta \in \mathbb{Q}(\alpha)$ as polynomial of degree at most $d - 1$ with integer coefficients (this is wlog since we can always clear denominators).

The four elementary operations in $\mathbb{Q}(\alpha)$ are the usual operations with polynomials, however special care should be taken [29] if $A$ is not the minimal polynomial of $\alpha$, which is usually the case. The complexity of the four operations are the complexity of the corresponding polynomial operations. However we can improve a lot the practical complexity if we represent the elements of $\mathbb{Q}(\alpha)$, which are univariate polynomials, in the Horner's basis [7]. The latter approach allow us to argue that the complexity of the four operations is quasi-linear, hence optimal up to logarithmic factors.

The most difficult elementary operation is the determination of the sign of $\beta \in \mathbb{Q}(\alpha)$, which also corresponds to the comparison of two numbers that belong to the same extension field. Since $\beta$ is represented by a polynomial $P$, it suffices to compute $\text{sign}(P(\alpha))$. This can be done easily using the results of Lem. 24.

**Remark 29** *Currently we are implementing an `Algebraic_extension` class in SYNAPS [30]. This class allows us to perform real root isolation and other elementary operations with polynomials that belong to $\mathbb{Q}(\alpha)[X]$. It is not clear at all, at least from a practical point of view, whether the computations of the signs of various quantities that are needed should be performed using Sturm-Habicht sequences or repeated refinements of the isolating intervals. We believe that a combination of both approaches will eventually lead to an optimal, from an implementation point of view, scheme.*

**Remark 30** *Computationaly, is very costly to perform operations with two numbers that belong to two different extension fields, or to change the extension field of a number. Currently only the package of Rioboo in AXIOM [34] can perform non trivial operations with real algebraic numbers that belong to different extension fields (actually to towers of extension fields).*

*However given an extension field $\mathbb{Q}(\alpha)$ we can easily compute the representations of the $\mathbb{Q}(-\alpha)$, $\mathbb{Q}(\frac{1}{\alpha})$, $\mathbb{Q}(a \pm \alpha)$, $\mathbb{Q}(a\,\alpha)$ and $\mathbb{Q}(\frac{\alpha}{a})$, $a \in \mathbb{Q}$, where by representation we mean the isolating interval representation of the real algebraic number that defines the extension field.*

*We will present the full details of algorithmic, complexity and implementation issues for computation in an extension filed(s) in a future report since this is work in progress.*

**Remark 31** *Recently we have implemented in* MAPLE *a prototype library, that besides real root isolation and sign evaluations also provides the four operations between real algebraic numbers that belong to different extension fields as well as fractional powers of real algebraic numbers. In the near future we will make the library freely available.*

**Remark 32** *As for* C++ *implementations, with freely available code, that can perform the basic operations with real algebraic numbers, i.e real root isolation and sign evaluations, the reader may refer to* SYNAPS *[30], where also the bivariate problems of the next section are treated, or to the library of Guibas et al [24], especially optimized for kinetic data structures, or to* NiX *the polynomial library of* EXACUS *[4], which is part of a bigger library for non linear computational geometry.*

# 4 Computations with 2 real algebraic numbers

In this section we will present some algorithmic and complexity results concering computations with two real algebraic numbers and real solving of bivariate polynomial systems.

The algorithms, the implementation details as well as experimental results for bivariate polynomial system solving were presented in [18]. In this work we present the complexity analysis.

## 4.1 Sturm-Habicht sequences for bivariate polynomials

**Theorem 33** *[33] Let* $F, G \in (\mathbb{Z}[Y_1, \ldots, Y_t])[X]$, *where* $\deg_X(F) = p \geq q = \deg_X(G)$ *and* $\mathcal{L}(F) = \mathcal{L}(G) = \tau$. *Moreover, let* $\deg_{Y_i}(F) \leq \delta_i$ *and* $\deg_{Y_i}(G) \leq \delta_i$, *where* $1 \leq i \leq t$. *We can compute the quotient boot of* $F$ *and* $G$, *the resultant and the gcd in*

$$\mathcal{O}_B\left(q \lg q \, \mathsf{M}\big((p+q)^{t+1}\delta_1 \cdots \delta_t \, \tau\big)\right)$$

**Remark 34** *Th. 33 generalizes Th. 4. It is easy to deduce this complexity from the univariate case if we transform the multivariate polynomials to univariate ones, using iterated applications of univariate "binary segmentation" and Kronecker's map [5, 40]. In an expanded version of this paper we will present complexity results for multivariate Sturm-Habicht sequences based on a more sophisticated binary segmentation [25] that can save some logarithmic factors and that has a simpler encoding and decoding algorithm, that involves only shifts and additions.*

**Remark 35** *If* $F$ *and* $G$ *are bivariate polynomials the complexity of computing the quotient boot, the resultant and the gcd is* $\mathcal{O}_B(q \lg q \, \mathsf{M}\left(\tau(p+q)^2\delta_1\right)) = \widetilde{\mathcal{O}}_B(qp^2\delta_1\tau)$. *The bit size of the gcd as well as the bit size of the polynomials in the quotient boot is bounded by* $\mathcal{O}(p\tau)$ *[2, 33, 40].*

The fact that Sturm-Habicht sequences are amenable to any specialization of the coefficients [21, 22] finds application when we are computing with multivariate (and in our case bivariate) polynomials.

Let $F$ and $G$ be two polynomials with parametric coefficients, such that their degree does not change after any specialization in the parameters. The computation of their Sturm-Habicht sequence before specialization of their coefficients, guarantees that the seuenece is valid under every specialization. We use this property so as to compute such a sequence for bivariate polynomials, regarding them either as polynomials in $(\mathbb{Z}[X])[Y]$ or in $(\mathbb{Z}[Y])[X]$. Remember that the last polynomial in the sequence is the resultant with respect to X or Y, respectively.

The following theorem will be very useful

**Theorem 36** *[2, 20, 23] Let* $f, g$ *square-free and coprime polynomials, such that* $\mathcal{C}_f$ *and* $\mathcal{C}_g$ *are in generic position. If*

$$H_j(X, Y) = \mathbf{StHa}_j(f, g) = h_j(X)Y^j + h_{j,j-1}(X)Y^{j-1} + \cdots + h_{j,0}(X)$$

*then if* $\zeta = (\alpha, \beta) \in \mathcal{C}_f \cap \mathcal{C}_g$ *then there exists* $k$, *such that*

$$h_0(\alpha) = \cdots = h_{k-1}(\alpha) = 0, \quad h_k(\alpha) \neq 0, \quad \beta = -\frac{1}{k}\frac{h_{k,k-1}(\alpha)}{h_k(\alpha)}$$

## 4.2 Bivariate sign evaluation

The previous tools suffice to compute the sign of a bivariate polynomial function evaluated over two algebraic numbers. Consider $F \in \mathbb{Z}[X, Y]$ and $\alpha \cong (A(x), I_1)$ and $\beta \cong (B(X), I_2)$ where $I_1 = [a_1, b_1]$, $I_2 = [a_2, b_2]$. We wish to compute the sign of $F(\alpha, \beta)$.

**Theorem 37 (Bivariate sign_at)** *Let* $F \in \mathbb{Z}[X, Y]$ *with* $\deg_X(F) = \deg_Y(F) = d_1$ *and* $\mathcal{L}(F) = \tau$ *and two real algebraic numbers* $\alpha \cong (A, I_\alpha), \beta \cong (B, I_\beta)$ *where* $A, B \in \mathbb{Z}[X]$, $\deg(A) = \deg(B) = d_2$, $\mathcal{L}(A) = \mathcal{L}(B) = \tau$ *and* $I_\alpha, I_\beta \in \mathbb{Q}^2$ *with bit sizes bounded by* $\mathcal{O}(d_2 \tau)$.

*We can compute the sign of* $F$ *evaluated over* $\alpha$ *and* $\beta$, *i.e* $\operatorname{sign}(F(\alpha, \beta))$ *in* $\widetilde{\mathcal{O}}_B(d_1^3 d_2^3 \tau)$ *time, assuming* $d_1 \leq d_2$.

**Proof.** In order to compute the sign of $F(\alpha, \beta)$, we consider $F$ as a univariate polynomial in $X$, i.e $F \in (\mathbb{Z}[Y])[X]$ and we try to compute its sign when we evaluate it over $\alpha$, as we did in the univariate case (Cor. 24).

We compute the Sturm-Habicht quotient boot of $A$ and $F$ with respect to $X$ in $\widetilde{\mathcal{O}}_B(d_1^2 d_2^2 \tau)$, as follows:

We transform $F$ to a univariate polynomial in $\mathbb{Z}[X]$, using binary segmentation induced by the map $\varphi : Y \mapsto 2^{c \, d_2 \tau}$ [25, 33, 40], where $c$ is a suitable constant. Now $\mathcal{L}(F) = \mathcal{O}(d_1 d_2 \tau)$. Thus, the computation of **StHaQ**$(A, F)$ takes $\widetilde{\mathcal{O}}_B(d_1^2 d_2^2 \tau)$ time, using Th. 3, since $p = d_1$, $q = d_2$ and the bit size is $\mathcal{O}(d_1 d_2 \tau)$ and dominates the cost of binary segmentation.

Notice that there are $\mathcal{O}(d_1)$ polynomials, in $\mathbb{Z}[X]$, in the sequence.

We evaluate the sequence on the left endpoint of $I_\alpha$, which has bit size $\mathcal{O}(d_2 \tau)$. The evaluation of the sequence can be done in $\widetilde{\mathcal{O}}_B(d_1 d_2^2 \tau)$, using Th. 5, where $p = d_2$, $q = d_1$, $\sigma = \mathcal{O}(d_2 \tau)$ and the bit size is $\mathcal{O}(d_1 d_2 \tau)$.

By applying the inverse transformation of $\varphi$ in the evaluated sequence we get $\mathcal{O}(d_1)$ univariate polynomials with respect to $Y$, with degrees $\mathcal{O}(d_1 d_2)$ and bit size $\mathcal{O}(d_1 d_2 \tau)$. For every such polynomial we compute its sign when we evaluate it over $\beta$. This can be done in $\widetilde{\mathcal{O}}_B(d_1^2 d_2^3 \tau)$ (Th. 5), which dominates the cost of the inverse transformation. Since there are $\mathcal{O}(d_1)$ polynomials, this step can be done in $\widetilde{\mathcal{O}}_B(d_1^3 d_2^3 \tau)$.

We do the same for the right endpoint of $I_\alpha$. This suffices to compute $\operatorname{sign}(F(\alpha, \beta))$ using Th. 10 or Cor. 24.

Thus the overall complexity of the operation is $\widetilde{\mathcal{O}}_B(d_1^3 d_2^3 \tau)$. □

**Remark 38** *The complexity of computing* **StHaQ**$(A, F)$ *can also be derived from Rem. 35, where* $p = d_2, q = \delta_1 = d_1$.

**Remark 39** *We can extend this approach to polynomials with an arbitrary number of variables, similar to [37]. However the usage of Sturm-Habicht sequences, instead of generalized Sturm sequences, improves both the theoretical [2] and the practical complexity [12, 18, 42].*

## 4.3 Two variants of bivariate real solving

In what follows we will present two variants of bivariate polynomial real solving. We assume that the polynomials are square-free. However we can easily drop this assumption.

### 4.3.1 Modified RUR

We use Th.36, following [23, 20], so as to compute the solution of bivariate polynomial systems. We consider polynomials $f, g \in \mathbb{Q}[X, Y]$, such that $\mathcal{C}_f, \mathcal{C}_g$ are in generic position and we compute the resultant of $f, g$ with respect to $Y$, which is a polynomial in $X$. The real solutions of the polynomial correspond to the $x-$ coordinates of the solution of the system. Then, using Th. 36, we lift these solutions in order to determine the $y-$coordinates, as a rational univariate function evaluated over an algebraic number. Even though the previous approach is straightforward, it has one main disadvantage. The y-coordinates are computed implicitly. If this is all that we want then this is not a problem. However in most cases we want to further manipulate the solutions of the system, i.e. to compare two y−coordinates or to count the number of branches of each curve above or below this ordinate. Of course we can always find the minimal polynomial of these algebraic numbers, but this is quite expensive. Thus we chose an alternatively way.

We compute the resultant, using the Sturm-Habicht Sequence, both with respect to Y and X, $R_x$ and $R_y$ respectively. We compute the isolating interval representation of the real roots of $R_x$ and $R_y$ (Th. 3.1) Let $\alpha_1 < \cdots < \alpha_k$ and $\beta_1 < \cdots < \beta_l$ be the real roots of $R_x$ and $R_y$, respectively. For the real roots of $R_y$ we compute rational intermediate points, $q_0 < \beta_1 < q_1 < \cdots < q_{l-1} < \beta_l < q_l$ where $q_j \in \mathbb{Q}, 0 \leq j \leq l$. We can easily compute the intermediate points, since the algebraic numbers are in isolating interval representation.

For every root $\alpha_i, 1 \leq i \leq l$, using Th. 36, we compute a rational univariate representation of the corresponding y-coordinate, which is without loss of generality, of the form $\gamma_i = \frac{A(\alpha_i)}{B(\alpha_i)}$. Since have already computed the real solutions of $R_y$, it suffices to determine to which $\beta_j$, $\gamma_i$ equals to, that is to find an index $j$ such that

$$q_j < \frac{A(\alpha_i)}{B(\alpha_i)} < q_{j+1}$$

or, if we assume that $A(\alpha_i) > 0$, this can be checked using Lemma 24, then

$$q_j A(\alpha_i) < B(\alpha_i) < q_{j+1} A(\alpha_i)$$

Actually what we really want is to determine the sign of univariate polynomials of the form $U(X) = q_j A(X) - B(X)$ evaluated over the real algebraic numbers that are solutions of $R_x = 0$. This can be done using Lemma 24.

**Remark 40** *We computed the complexity of the modified RUR algorithm is $\widetilde{\mathcal{O}}_B(d^{10}\tau^2)$, since the complexity of the algorithm is dominated by the real solving of the univariate solutions. However we will present the complexity analysis in an extended version of the paper, since we are currently working on the complexity of transforming two algebraic curves to generic position and on the complexity of transforming the real solutions back to the original coordinate system if a random shear is performed as well as on the complexity of the topology of real algebraic curves (see also Rem. 41).*

**Remark 41 (Topology of an algebraic curve in 2D)** *As stated by G. Vega and El Kahoui [20] the bottleneck of the algorithm for the topology of a plane real algebraic curve is the computation of the Thom's codes of the real roots of a univariate polynomial. In our notation this is $\widetilde{\mathcal{O}}_B(d^{14}\tau^2)$.*

*The complexity results that we presented for univariate real root isolation (Th. 19) can improve their bound. We claim that the bound for the topology algorithm is $\widetilde{\mathcal{O}}_B(d^{10}\tau^2)$, induced by the modified RUR algorithm, even when the curves are not in generic position. However, for this very crucial problem in non-linear computational geometry a more detailed analysis is needed so as to give light to various details of the algorithm in order to present an efficient implementation for real-life applications.*

### 4.3.2 Naive real solving

In many cases, especially in non-linear computational geometry or when we want to solve a system of inequalities, it is very difficult to deal with bivariate polynomials that are not in generic position.

Even though the assumption of generic position is without loss of generality since we can apply a transformation of the form $(X, Y) \mapsto (X + aY, Y)$, where $a$ is either a random number or a number computed deterministically [20] before the execution of the algorithm, or we can detect non-generic position during the execution [23, 2], then apply a transformation of the form $(X, Y) \mapsto (X + Y, Y)$ and start the algorithm recursively. However neither approach is an easy computational task. Moreover if we need the solution in the original coordinate system then we must perform the inverve transformation, and this is a very difficult task, especially when we want to avoid refinements that can go up to seperation bounds.

In order to overcome such barriers we suggest one (naive) variant for bivariate polynomial system solving. The intersting reader may refer to [18] for a complete descritption of the algorithm and for preliminary experimental results.

**Theorem 42 (Naive solve)** *Let* $F, G \in \mathbb{Z}[X, Y]$ *be square-free polynomials with total degree bounded by* $d$ *and (coefficient) bit size bounded by* $\tau$. *We can compute the real solutions of the system* $F = G = 0$ *in* $\widetilde{\mathcal{O}}_B(d^{14}\tau)$, *under the assumption that* $\tau = \widetilde{\mathcal{O}}_B(d^4)$.

**Proof.** We compute the resultant of $F$ and $G$ with respect to $Y$, i.e $R_X$ in time $\widetilde{\mathcal{O}}_B(d^4\tau)$ (Rem. 35). Note that $\deg(R_x) = \deg(R_Y) = \mathcal{O}(d^2)$ and $\mathcal{L}(R_X) = \mathcal{L}(R_Y) = \mathcal{O}(d\tau)$.

We compute the real algebraic numbers that are roots of $R_X$, in isolating interval representation in $\widetilde{\mathcal{O}}_B(d^{10}\tau^2)$ time (Th. 19). Their representation involves the square-free part of $R_X$, also of coefficient bit-size $\mathcal{O}(d\tau)$ (Th. 8), and an isolating interval, whose endpoints have bit-size $\mathcal{O}(d^3 \tau)$ (Rem. 20). There are at most $\mathcal{O}(d^2)$ real roots, i.e $\deg(R_X)$. We do the same with respect to $Y$.

We form all the possible pairs of real algebraic numbers and for every pair of algebraic numbers we test if both $F$ and $G$ are zero. There are $\mathcal{O}(d^4)$ such tests and each takes $\widetilde{\mathcal{O}}_B(d^{10}\tau)$ time (Th. 37).

Thus the overall complexity is $\widetilde{\mathcal{O}}_B(d^{14}\tau)$.  □

## 4.4 Bivariate simultaneous inequalities

Let $P, Q, A_1, \ldots, A_{n_1}, B_1, \ldots, B_{n_2}, C_1, \ldots, C_{n_3} \in \mathbb{Z}[X, Y]$, with degree bounded by $d$ and coefficient bit size bounded by $\tau$. We wish to compute the number and the real roots, $(\alpha, \beta)$, of the system $P(X, Y) = Q(X, Y) = 0$ such that $A_i(\alpha, \beta) > 0$, $B_j(\alpha, \beta) < 0$ and $C_k(\alpha, \beta) = 0$ and $1 \leq i \leq n_1, 1 \leq j \leq n_2, 1 \leq k \leq n_3$. Let $n = n_1 + n_2 + n_3$.

**Corollary 43** *There is an algorithm that solves the problem of simultaneous inequalities in time* $\widetilde{\mathcal{O}}_B(\max\{nd^{12}\tau, d^{14}\tau\}) = \widetilde{\mathcal{O}}_B(d^{12}\tau \max\{n, d^2\})$.

**Proof.** First we compute the isolating interval representation of all the real roots of the system $P = Q = 0$ in $\widetilde{\mathcal{O}}_B(d^{14}\tau)$ (Th. 42). There are at most $d^2$ real roots of the system. For every real root $(\alpha, \beta)$ of $P$, for every polynomial $A_i$, $B_j$, $C_k$ we compute the sign $A_i(\alpha, \beta)$, sign $(B_j(\alpha, \beta))$ and sign $(C_i(\alpha, \beta))$.

Sign determination costs $\widetilde{\mathcal{O}}_B(d^{10}\tau)$ (Th. 37) and in the worst case we must compute $d^2$ of them. Thus the overall cost is $\widetilde{\mathcal{O}}_B(\max\{nd^{12}\tau, d^{14}\tau\})$.  □

# 5 Conclusions and future work

We plan to apply our tools in computing the topology of algebraic curves in 2D and 3D, as well as the topology of surfaces in 3D. Another possible approach, to be implemented and compared at a practical ant theoritical level, includes the adoption and Thom's encoding [2, 10, 36]. Last but not least, we intend to use arithmetic filtering to handle cases that are far from degenerate, so as to improve the speed of our software for generic inputs.

**Acknowledgements:** The second author acknowledge inspirational comments by Bernard Mourrain and especially his indication of using Horner's basis for computations in an extension field. Both authors acknowledge partial support by INRIA's associated team project "CALAMATA", a bilateral collaboration between the GALAAD group of INRIA Sophia-Antipolis (France) and the National Kapodistrian University of Athens, and IST Programme of the EU as a Shared-cost RTD (FET Open) Project under Contract No IST-006413-2 (ACS - Algorithms for Complex Shapes).